\documentstyle[aps,prl,twocolumn,epsf,epsfig]{revtex}
\def\beq{\begin{equation}}
\def\eeq{\end{equation}}
\def\beqarr{\begin{eqnarray}}
\def\eeqarr{\end{eqnarray}}

\begin{document}
\draft
\twocolumn[\hsize\textwidth\columnwidth\hsize\csname @twocolumnfalse\endcsname

\title{Quasi-Two Dimensional Diluted Magnetic Semiconductor Systems}
\author{D. J. Priour, Jr, E. H. Hwang, and S. Das Sarma}
\address{Condensed Matter Theory Center, Department of Physics,
University of Maryland, College Park, MD 20742-4111}


\maketitle

\begin{abstract}
We develop a theory for two-dimensional diluted magnetic semiconductor 
systems (e.g. $\textrm{Ga}_{1-x}\textrm{Mn}_{x}\textrm{As}$ layers) 
where the itinerant carriers
mediating the ferromagnetic interaction between the impurity local 
moments, as well as the local moments themselves, are confined in a 
two-dimensional layer.  The theory includes exact spatial disorder 
effects associated with the random local moment positions within a 
disordered RKKY lattice field theory description.  We predict the 
ferromagnetic transition temperature ($T_{c}$) as well as the nature 
of the spontaneous magnetization.  The theory includes disorder and 
finite carrier mean free path effects as well as the important 
correction arising from the {\it finite temperature} RKKY 
interaction, finding a strong density dependence of $T_{c}$ in contrast
to the simple virtual crystal approximation.  
\end{abstract}

]

Many projected applications of diluted magnetic semiconductors (DMS), i.e. 
systems which combine the advantages of a ferromagnetic material with those
of a semiconductor with the additional flexibility of carrier-mediated 
ferromagnetism enabling the tuning of the magnetic properties by applying
external gate voltages or optical pulses to control the carrier density, 
would involve the use of two-dimensional (2D) DMS structures such as quantum 
wells, multilayers, superlattices, or heterostructures.  Such 2D DMS
structures 
are also of intrinsic fundamental interest since magnetic properties in 
two dimensions are expected to be substantially different from the three 
dimensional (3D) systems \cite{dietl} that have mostly been
theoretically studied in the  
DMS literature.  The 2D DMS systems introduce the possibility of gating, and
thereby controlling both electrical and magnetic properties by tuning the 
carrier density.  In fact, such a carrier density modulation of DMS
properties 
has already been demonstrated~\cite{uno,dos} in gated DMS field effect
heterostructures. 
For various future spintronic applications the development of such 2D
DMS structures is obviously of great importance.

     In this Letter, we provide the basic theoretical picture underlying
2D DMS ferromagnetism focusing on the well-studied $\textrm{Ga}_{1-x}
\textrm{Mn}_{x}\textrm{As}$, with $x \approx 0.03 - 0.08$, a system
where the  
ferromagnetism is well-established to be arising from the alignment (for
$T < T_{c} \sim 100K-200K$) of Mn local moments through the indirect exchange
interaction carried by itinerant holes in the GaAs valence (or impurity) 
band (that are also contributed by the Mn atoms which serve the dual purpose
of being the impurity local moments as well as the dopants).  Our theory is 
quite general and should apply to other ``metallic'' DMS materials where the
ferromagnetic interaction between the impurity local moments is mediated by 
itinerant carriers (electrons or holes).  Our theory is a suitable 2D 
generalization of both the continuum lattice version of the highly successful 
RKKY mean field theory (MFT) that has earlier been applied \cite{dietl} 
to 3D metallic $\textrm{Ga}_{1-x}\textrm{Mn}_{x}\textrm{As}$ systems.  One 
of our important findings is that the ferromagnetic transition (``Curie'') 
temperature for the 2D DMS systems typically tends to be substantially less
than the corresponding 3D case with equivalent system parameters.  In 
particular, $T_{c}$ in the 2D case is found to be comparable
($T_{c} \sim T_{F}$)  
to the Fermi temperature ($T_{F} = E_{F}/k_{B}$, where $E_{F}$ is the 2D Fermi 
energy) of the 2D hole systems.  This necessitates that the full finite 
temperature form for the 2D RKKY interaction be used in calculating
DMS magnetic  
properties, further suppressing $T_{c}$ in the system.  In fact, this general 
lowering of the 2D DMS $T_{c}$ compared with the corresponding 3D case
is our central  
new theoretical result.  This implies that spintronic applications
involving 2D  
DMS heterostructures will be problematic since the typical $T_{c}$ (at
least for  
the currently existing DMS materials) is likely to be far below room
temperature 
($T_{c} < 100K$).  A related equally important theoretical finding is that, 
although the continuum virtual crystal mean field approximation much
used for 3D  
$\textrm{Ga}_{1-x}\textrm{Mn}_{x}\textrm{As}$ physics~\cite{cinco} 
would predict that the $T_{c}$ in 2D DMS systems
(being proportional to the 2D density of states) is independent of 2D carrier
density, there is an {\it intrinsic} carrier density dependence of 
$T_{c}$ even in the strict 2D system arising from the density and temperature
(i.e. $T/T_{F}$) dependence of the finite-temperature effective RKKY
interaction. 

We specifically consider the so-called delta-doping schemes for 
localizing Mn dopants, the species responsible for DMS ferromagnetism,
to very narrow layers~\cite{dos}.  It is this situation which we examine in 
this letter, 
spin-5/2 Mn impurities confined to a single plane in GaAs; although
the experimental reality may well be one in which a significant number
of impurities lie off this plane and/or the Mn occupy a different 
crystallographic plane, our results are qualitatively 
correct in these cases as well.  

\begin{figure}
\centerline{\psfig{figure=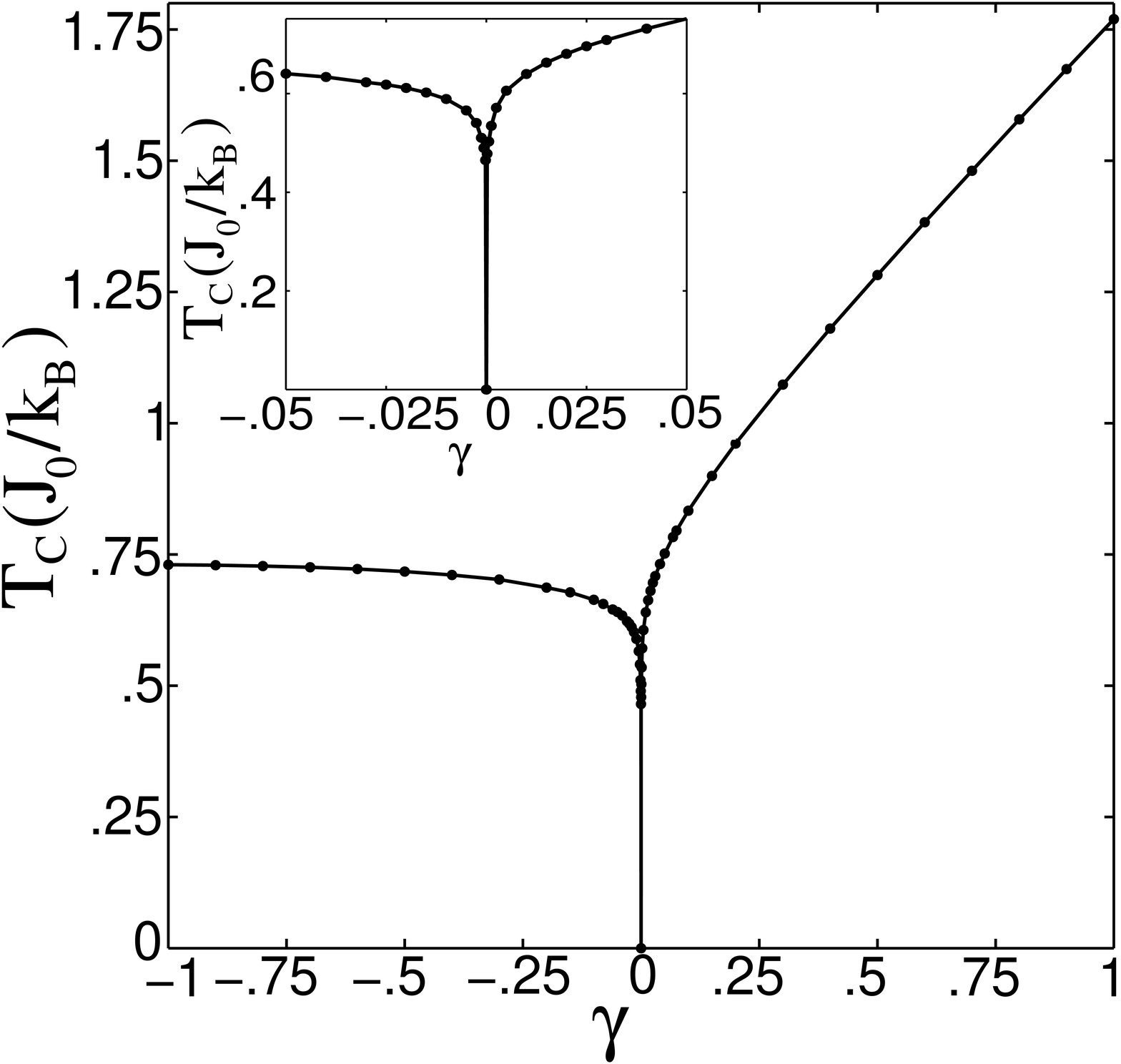,width=2.75in}}
\caption{Curie Temperatures plotted versus $\gamma$, the 
anisotropy parameter.  The inset displays $T_{c}$ values for 
a much smaller range of anisotropies.  In both images, the Monte
Carlo errors are smaller than the graph symbols.
The black line is included only as a guide
to the eye in both the main graph as well as the inset.}
\label{Fig:fig-2}
\end{figure}

Although the Hohenberg-Mermin-Wagner theorem precludes
long range order in 2D systems with Heisenberg spins, 
the theorem applies only for the case in which the coupling
between spins is absolutely isotropic.  As has been shown
both formally and numerically~\cite{cincodemayo,guerradelasestrellas,qmc},
even a small amount
of anisotropy is sufficient to stabilize long range order at 
finite temperatures.  We examine the impact of anisotropy 
explicitly by studying the classical anisotropic 2D Heisenberg
model ${\mathcal H} = 
-J_{0} \sum_{\langle ij \rangle}[S^{x}_{i}S^{y}_{j} + S^{x}_{i}S^{y}_{j} +
(\gamma + 1)S^{z}_{i}S^{z}_{j}]$, where $J_{0}$ is the ferromagnetic exchange 
constant, $\gamma$ is the anisotropy parameter, and the spins 
occupy a 2D square lattice; the sum is over nearest-neighbors only.
Using a variant of the Wolff cluster Monte Carlo technique 
(with modifications appropriate
for a finite $\gamma$~\cite{thalia}), we calculate Curie temperatures by finding the 
intersection of 
Binder cumulants (given by $U = 1 - \frac{1}{3}\langle|M|^{4} \rangle/\langle |M|^{2}
\rangle$~\cite{ciara}) for two different system sizes ($L = 40$ and $L = 80$).  The results
are shown in Fig.~\ref{Fig:fig-2}, where it is readily evident from the main 
graph and the inset (depicting $T_{c}$ for a smaller range of $\gamma$ values)
that even a small amount of anisotropy yields Curie temperatures in the 
vicinity of $J_{0}/k_{B}$, a value in reasonable agreement with 
$T_{c}^{MFT} = \frac{4}{3}J_{0}/k_{B}$, the corresponding mean field 
result for the nearest neighbor 2D Heisenberg model.  Previous
theoretical studies also examined ferromagnetism in the 2D DMS 
context~\cite{trespuntocinco,trei}, without, however, considering the 
Hohenberg-Mermin-Wagner theorem or the anisotropy issue.

We apply the lattice MFT 
developed previously~\cite{tres}, but we also find intriguing 
results in the context of simple continuum Weiss MFT
which demonstrate that it is essential to incorporate finite temperature
effects in the effective carrier-mediated interaction between Mn
impurity moments. 
We assume the 2D hole gas to be confined in the same plane as the Mn dopants--
it is straightforward to consider~\cite{trespuntocinco,trei}
a spatial separation between the dopants
and the holes as well as to consider the quasi-2D confinement for the holes.
These additional complications would lower $T_{c}$ below the strict 2D limit 
considered in our model.  

Our theory is constructed for two dimensional DMS systems in the 
metallic limit with itinerant carriers (we assume the carrier-mediated
effective Mn-Mn indirect exchange interaction to be of the RKKY 
form).  However, including a finite carrier mean free path in 
our theory allows us to take into account the dependence of the 
magnetic behavior of our system on the carrier transport properties.
In fact, using an exponential cutoff in the range of the RKKY function 
permits us to treat the long and short-range magnetic interaction regimes 
simply by varying the cutoff parameter $l$.
In our system, salient parameters include the Mn local moment concentration
($x$), the free carrier density ($n_{c}$), and the exponential cutoff scale $l$
associated with the carrier mean free path.  The 2D Fermi temperature $T_{F} = 
\frac{\hbar^{2}k_{F}^{2}}{2m^{*}k_{B}}$ depends on $n_{c}$ since the 2D
$k_{F} \propto n_{c}^{1/2}$,
where $k_{F}$ is the Fermi wave vector.  We take the carrier effective
mass to be  
$m^{*} = 0.4m_{e}$ for all our numerical results.  All quantities
($T_{F}, k_{F}, r$, etc.) 
appearing in our theory refer to 2D parameters.  

\begin{figure}
\centerline{\psfig{figure=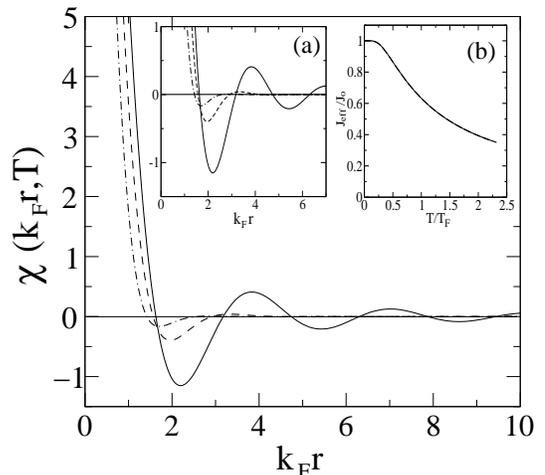,width=2.75in}}
\caption{Temperature dependent 2D RKKY range function $\chi(k_{F}r,T/T_{F})$
as a function of $k_{F}r$ for several temperatures 
$T/T_{F} = 0.0, 0.5$, and $1.0$ (corresponding to the solid, dashed,
and dot dashed lines respectively).  Inset (a) also shows
$\chi(k_{F}r,T/T_{F})$, and inset (b) portrays the effective finite
temperature coupling $J_{\textrm{eff}}$.}
\label{Fig:fig-1}
\end{figure}
 
Our effective Hamiltonian describes the Mn-Mn magnetic interaction 
between classical Heisenberg spins ${\bf S}_{i}$ on a 2D lattice:
\begin{equation}
{\mathcal H} = \sum_{ij} J^{RKKY}_{ij} {\bf S}_{i} \cdot {\bf S}_{j} 
\label{Eq:eq1}
\end{equation}
where ${\bf S}_{i}$ is the $i$th Mn local moment of spin $5/2$. 
In our lattice MFT we assume the Mn dopants
to lie entirely on the [100] plane of  
the GaAs zinc-blende crystal lattice, with the impurities occupying the 
Ga sites, which form a square lattice with lattice constant $a$.

     In our Hamiltonian, the carrier mediated RKKY
indirect exchange interaction describes the effective magnetic interaction 
between Mn local moments induced by the free carrier spin 
polarization.  $J^{RKKY}_{ij}(T) \equiv {\mathcal
  J}_{0}\chi(k_{F}r,T/T_{F})$, where 
$\chi(k_{F}r,T/T_{F})$ is the temperature dependent range function which is
obtained from the 2D spin susceptibility, and ${\mathcal J}_{0} \equiv 
\left[\frac{J_{pd}}{a} \right]^{2} \frac{m^{*}}{8[\pi
  \hbar]^{2}a^{2}}$.  Taking  
$J_{pd}=0.15~\textrm{eV} \textrm{nm}^{3}$~\cite{cuatro} and square lattice
constant $a=.4\textrm{nm}$, we find 
${\mathcal J}_{0} = 0.10~\textrm{eV}$; we use this value for ${\mathcal J}_{0}$
throughout this work.

For $T=0$ the 2D RKKY interaction is known exactly \cite{Aristov}:
\begin{equation}
J^{RKKY}_{ij}
\equiv {\mathcal J}_{0} (k_{F}a)^{2}\left[ J_0(k_Fr)N_0(k_Fr) +
  J_1(k_Fr)N_1(k_Fr) \right] 
\label{Eq:eq5}
\end{equation}
where $J_n(x)$ are the Bessel functions of the first kind, and $N_n(x)$
are Bessel functions of the second kind. The asymptotic form of the
RKKY interaction for $k_F r \gg 1$ is $J^{RKKY}(r) \propto {\mathcal J}_0
\sin(2k_Fr)/r^2$. 
At finite temperatures (i.e., for $T/T_{F} \ne 0$), 
it is not possible to obtain the 
RKKY range function analytically.  Thus, we calculate the finite
temperature RKKY interaction numerically.
In Fig.~\ref{Fig:fig-1}, we show the temperature dependent range 
function $\chi(k_{F}r,T/T_{F})$ as a function of $k_{F}r$  
for various 
$T/T_{F}$ values.   
One sees increasingly severe thermal damping of the RKKY oscillations with
increasing $T/T_{F}$.  Inset (a) of Fig.~\ref{Fig:fig-1} provides a closer
view of the RKKY oscillations depicted in the main graph, while 
inset (b) displays the effective coupling constant, given by 
$J_{\textrm{eff}} = \int J_{ij}^{RKKY}({\bf r}) d{\bf r}$.  One
finds, as expected, that $J_{\textrm{eff}}$ decreases with increasing
temperature due to the damping of the range function.  Since the 
magnetic properties of the 2D system are directly dependent on
$J_{\textrm{eff}}$, a finite $T/T_{F}$ can have an important impact. 

\begin{figure}
\centerline{\psfig{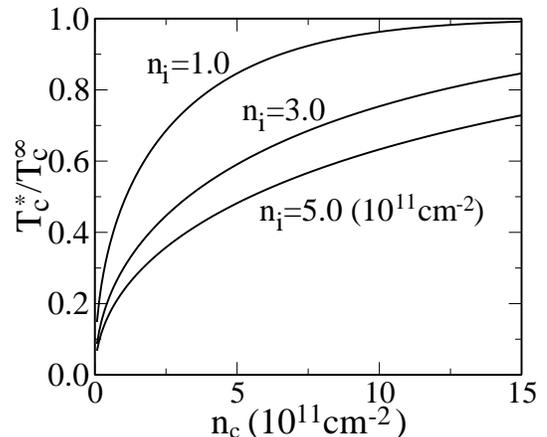}}
\caption{2D Curie temperature curves in the continuum 
MFT for various values of the magnetic impurity 
density $n_{i}$ ($n_{i} = xa^{-2}$).  Here $T_{c}^{\infty}$ 
is the continuum $T_{c}$ calculated
for $T_{F}=\infty$, i.e. the virtual crystal approximation result.}
\label{Fig:fig1}
\end{figure}

     In the lattice MFT (where we have in mind the 
previously mentioned square lattice with lattice constant $a$), we
calculate the Curie temperature with \cite{tres}
\begin{equation}
T_{c} = \frac{35}{12k_{B}}x \sum_{i=1}^{\infty}N_{i}J(r_{i},T_{c}),
\label{Eq:eq6}
\end{equation}
where $N_{i}$ and $r_{i}$ are the numbers and distances of the {\it i}th 
nearest neighbors, respectively.  
The continuum limit,
which we examine first, is attained 
for $l,k_{F}^{-1} \gg a$, where $a$ is the 
lattice spacing.  We examine the large $l$ limit ($l \gg k_{F}^{-1}$)
and find that   
Eq.~\ref{Eq:eq6} becomes 
\begin{eqnarray}
T_{c}^{*} &=& 
\frac{35\pi x{\mathcal J}_{0}}{6k_{B}}k_{F}^{2} \int_{0}^{\infty}  \chi
\left( k_{F}r,
T_{c}^{*}/T_{F} \right) r dr \nonumber \\
&=& \frac{35\pi x{\mathcal J}_{0}}{6k_{B}}g\left(T_{c}^{*}/T_{F}\right),
\label{Eq:eq7}
\end{eqnarray}
where $T_{c}^{*}$ is the Curie temperature in the continuum limit, and
$g(T_{c}^{*}/T_{F})$ depends only on the ratio of 
$T_{c}^{*}$ to the Fermi temperature
$T_{F}$.  For $T_{c}^{*} \ll T_{F}$, $g(T_{c}^{*}/T_{F}) \rightarrow 1$, 
and the dependence on carrier concentration is lost, and $T_{c}^{*}$ 
depends only on the impurity concentration $x$.  (For convenience, we
identify this temperature as
$T_{c}^{\infty} \equiv T_{c} (T_{F} \rightarrow \infty) = \frac{35\pi
  x {\mathcal J}_{0}}{6k_{B}}$.) 
This result, with no 
dependence on $n_{c}$, is proportional to the density of states at 
the Fermi Level $E_{F}$ in two
dimensions, and reflects the peculiarity of the 2D density of states
being independent of  
carrier density.  (The analogous continuum 3D
result, also proportional to the density of states at $E_{F}$, 
is $T_{c}^{*} \propto xn_{c}^{1/3}$, and {\it does} 
depend on the carrier density).  
For the more general case of finite $T_{c}^{*}/T_{F}$, one can readily 
calculate the Curie temperature 
self-consistently via numerical means.  Fig.~\ref{Fig:fig1} shows $T_{c}^{*}$
curves as a function of carrier density for various values of $x$, the 
Mn concentration.  A salient feature of the curves is a marked 
dependence on $n_{c}$, in contrast to what one finds when the (T=0)
RKKY formula is used (where there is no dependence at all on $n_{c}$).
Eventually, the curves saturate as $n_{c}$ is 
increased.  This is natural, since $T_{F}$ increases with $n_{c}$, and 
ultimately $T_{c}^{*}/T_{F}$ approaches zero.  This $T_{F} \rightarrow
\infty$     
limit of the continuum MFT is the actual virtual
approximation limit.

\begin{figure}
\centerline{\psfig{figure=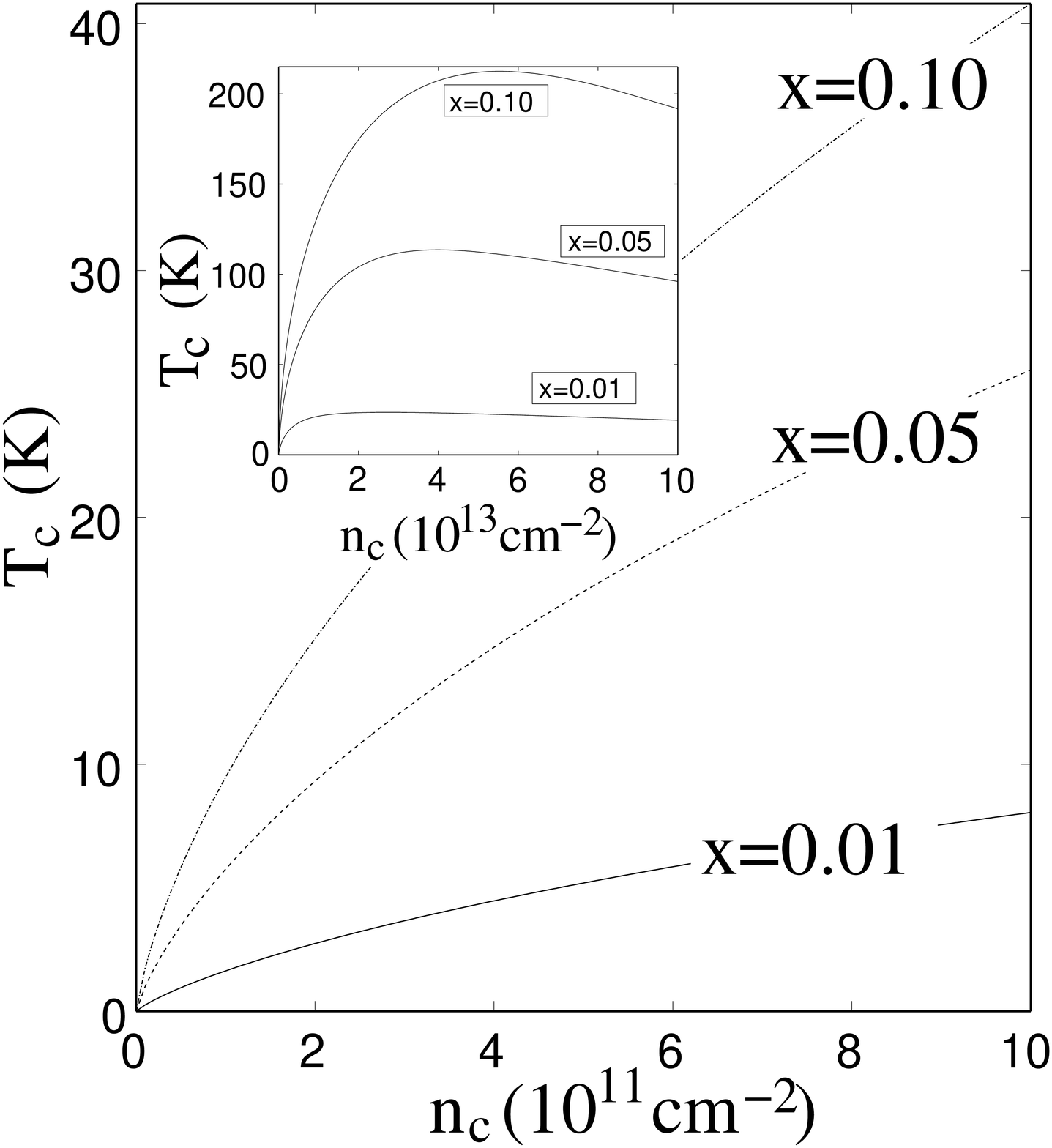,width=2.5in}}
\caption{Lattice MFT 2D 
Curie temperature curves for various values of $x$.  In
the inset, $T_{c}$ curves are shown for a much  
greater range of carrier density $n_{c}$. In both the main 
graph and the inset, $l/a = 5$.}
\label{Fig:fig2}
\end{figure}

Returning to the lattice MFT, which explicitly takes into 
account the discreteness of the square lattice, we also incorporate
the effects of the finite carrier mean free path $l$ by including an 
exponential cutoff in the range of the RKKY interaction. 
$T_{c}$ curves are shown in Fig.~\ref{Fig:fig2} for
several Mn dopant concentrations.
As in Fig.~\ref{Fig:fig1}, the Curie temperature increases monotonically
in $n_{c}$ over the experimentally accessible range of carrier 
densities.  However, for considerably higher $n_{c}$ (i.e. approaching
$10^{14}~\textrm{cm}^{-2}$), nonmonotonic behavior is seen in $T_{c}$.
This is evident in the inset of Fig.~\ref{Fig:fig2}, where the
Curie temperature curves seem to achieve saturation for 
intermediate carrier densities and ultimately  
begins to decrease as the length scale $k_{F}^{-1}$ of the 
RKKY oscillations shrinks relative to the lattice constant $a$.  Eventually, 
for $k_{F}^{-1} \sim a$, the discrete nature of the lattice sum in 
Eq.~\ref{Eq:eq6} has a strong effect, leading to a considerably
smaller $T_{c}$  
than the continuum value, $T_{c}^{*}$.  This result for 2D $T_{c}$ including 
disorder and finite temperature RKKY effects is one of our main new results.

\begin{figure}
\centerline{\psfig{figure=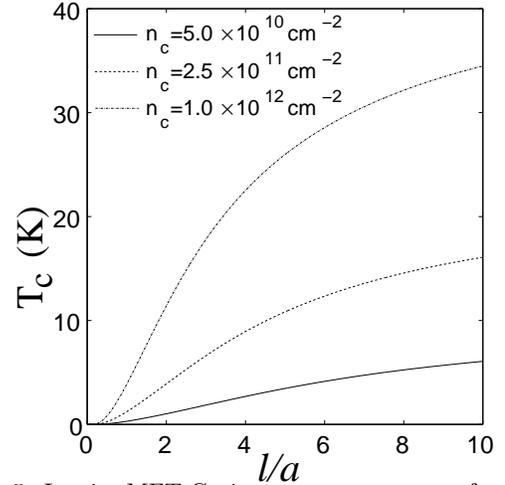,width=2.5in}}
\caption{Lattice MFT
Curie temperature curves for various values of $n_{c}$ plotted
as a function of the mean free path $l/a$.  In each case the 
same Mn concentration, $x=.05$, is used.}
\label{Fig:fig3}
\end{figure}

$T_c$ is strongly affected by the finite size of $l$;
in fact, one sees that for each of the carrier concentrations 
represented in Fig.~\ref{Fig:fig3}, $T_{c}$ initially increases sharply
with increasing $l$, eventually saturating for $l/a \gg 1$.
It is informative to examine the regime $a \ll l,k_{F}^{-1}$, where one can 
operate in the continuum limit. We introduce $\eta \equiv k_{F}l$ as an
important dimensionless variable; as will be seen, the ratio $T_{c}^{*}/T_{F}$
tends to zero as $\eta$ becomes small.
For $T_{c} \ll T_{F}$ and $k_{F}r \ll 1$, a 
reasonable approximation for 
the finite temperature RKKY range function is $J^{RKKY}(r,T)
\approx J^{RKKY}(r,0)[(1-a_{2}\tau^{2}-a_{3}\tau^{3}) 
+ (b_{2}\tau^{2} + b_{3}\tau^{3})(k_{F}r)]e^{-r/l}$ where $\tau \equiv
T/T_{F}$, 
$a_{2}=0.2153$, $a_{3}=-0.140$, $b_{2}=-1.333$, and $b_{3}=0.862$ 
(The $a_{i}$ and $b_{i}$ have been calculated numerically).
Since $J^{RKKY}(r,0)$ varies 
slowly with $k_{F}r$ for small $k_{F}$, we can replace the Bessel functions in 
Eq.~\ref{Eq:eq5} with a series expansions in $k_{F}r$, and the result for 
small values of this expansion variable is
\begin{eqnarray}
&J&^{RKKY}(r,T_{c}^{*}) \approx {\mathcal J}_{0}k_{F}^{2}/\pi 
\left[ \alpha_{1}  
+ \alpha_{2}(k_{F}r)^{2} \right]   \nonumber \\
&\times& \left[1 - (a_{2}\tau^{2} + a_{3}\tau^{3}) + (b_{2}\tau^{2} + b_{3}
\tau^{3})(k_{F}r) \right]e^{-r/l}, 
\label{Eq:eq9}
\end{eqnarray}
where $\alpha_{1} \equiv [-1/2 + \gamma + \ln(k_{F}r/2)]$,
and $\alpha_{2} \equiv [-3/16 + \gamma/4 + 1/4\ln(k_{F}r/2)]$; $\gamma = 
.57722 \ldots$ 
is 
the Euler constant.  In calculating $T_{c}^{*}$ (in the continuum case), 
we assume that although 
$l \ll k_{F}^{-1}$, $l \gg a$. 
This additional condition allows the replacement of 
the discrete formula for the Curie Temperature in Eq.~\ref{Eq:eq6} with the 
continuum version, and one has $T_{c}^{*} = 
x\frac{35\pi}{6}\int^{\infty}_{0}J^{RKKY}
(r,T_{c}^{*})rdr$
Carrying out the integration and solving for $T_{c}^{*}$, we find 
\begin{equation}
\label{Eq:eq10}
T_{c}^{*} = {\mathcal J}_{0}\frac{35x}{6k_{B}} 
\left\{ \eta^{2} \left[ \ln(\frac{2}{\eta}) - 1/2 \right]
+ \eta^{4} \left[-\frac{41}{16} +  \frac{1}{2}\ln(\frac{2}{\eta})
\right]\right \}, 
\end{equation}
where $\eta \equiv k_{F}l \ll 1$ and 
only terms up to fourth order in $\eta$ are shown.  The expression 
for $T_{c}^{*}$ given in Eq.~\ref{Eq:eq10} reveals a strong suppression of 
$T_{c}^{*}$ due to localization effects;  when $k_{F}$ is small
(i.e. in the small $n_{c}$ limit) the RKKY range function is very 
extended relative to $a$ and $l$. As a consequence, most of the 
RKKY interaction is 
truncated by the exponential cutoff associated with the finite mean free 
path $l$.  This truncation effect is so severe that $T_{c}^{*}$ is very 
small in comparison with $T_{F}$; to fourth order in $\eta$ there are 
no corrections to arising from the finiteness of $T_{c}^{*}/T_{F}$.
Note that in the opposite limit of $k_{F}l \gg 1$, which rarely applies
to DMS systems which are at best ``bad metals'', one can obtain the simple 
formula $T_{c}(k_{F}l \gg 1) \approx
T_{c}^{*}[1-(k_{F}l)^{-1}/\pi]$ by  
simply considering the disorder induced suppression of the 2D density of 
states.

\begin{figure}
\centerline{\psfig{figure=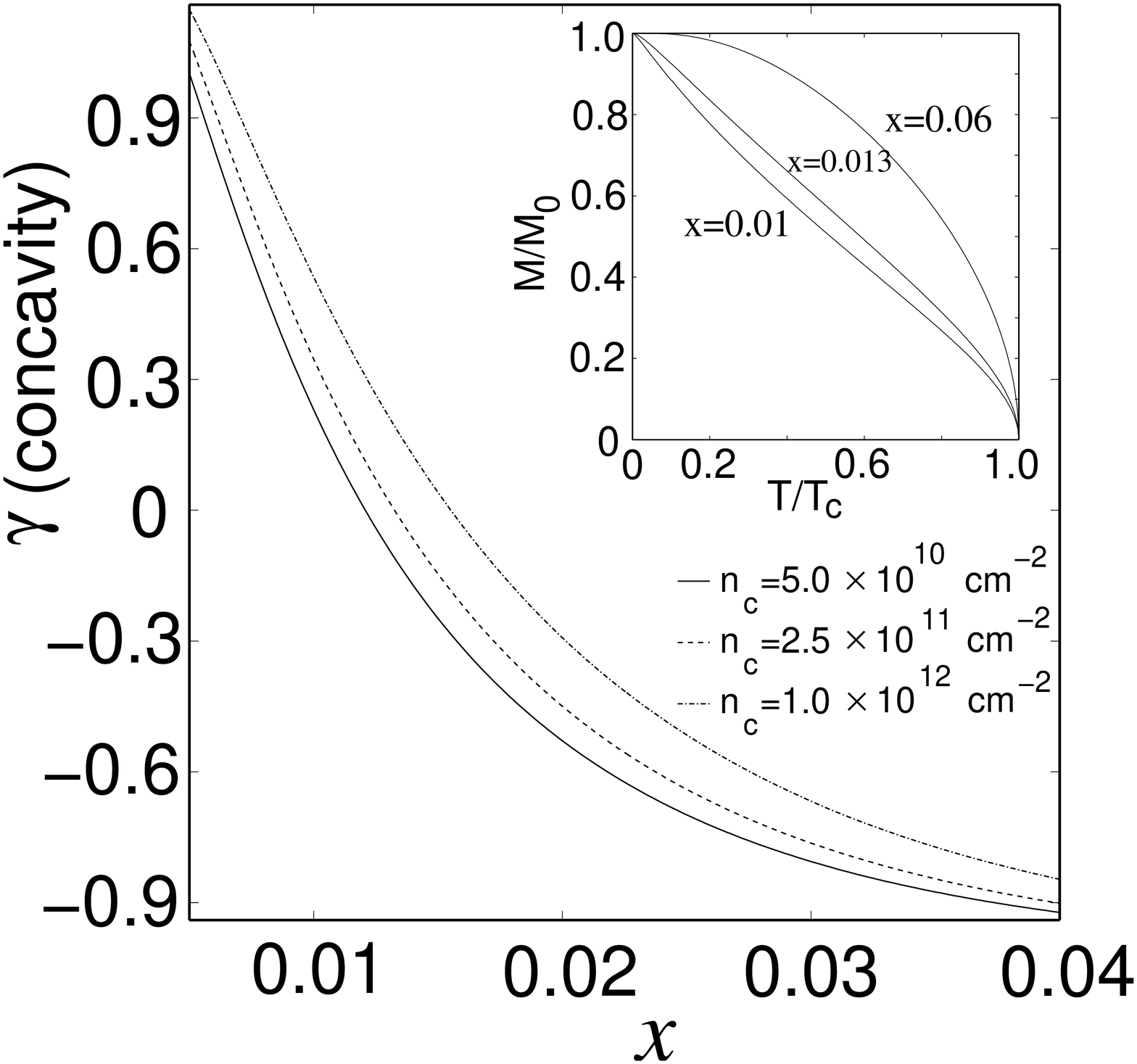,width=2.75in}}
\caption{Concavity curves versus Mn concentration $x$ for 
several carrier densities $n_{c}$ with $l/a=2.0$.  
The inset shows representative 
examples of the magnetization, $M(T)$.  All $M(T)$ curves are 
calculated for $n_{c}=2.5 \times 10^{11}~\textrm{cm}^{-2}$.}
\label{Fig:fig4}
\end{figure}

     In addition to $T_{c}$, one can also use lattice MFT
to calculate the magnetization $M(T)$~\cite{tres}.  The magnetization
behavior is influenced by the Mn impurity concentration as well as the 
form of the effective interaction between Mn local moments, in 
principle specified by the carrier mean free path $l$ and $n_{c}$.  For
convenience, we study $M(T/T_{c})$; using the normalized temperature 
scale $T/T_{c}$ allows a systematic comparison of magnetization 
profiles corresponding to different values of the parameters 
$l$, $n_{c}$, and $x$.  An important trait of the magnetization 
profile is its degree of {\it concavity}; concavity in $M(T)$ is a 
hallmark of an insulating system, while a convex profiles occur 
well within the metallic regime~\cite{cinco}.  Linear magnetization curves 
correspond to intermediate impurity densities and mean free paths.  
In terms of the magnetization, the concavity $\gamma$ is given by 
$\gamma \equiv \int^{t_{2}}_{t_{1}} M^{''}(T)dT$, or the difference
in the slopes of $M(T)$.  The sign of $\gamma$ indicates whether
$M(T)$ is convex (negative $\gamma$), concave (positive $\gamma$),
or linear (if $\gamma \approx 0$).  The temperatures $t_{1}$ and 
$t_{2}$ are selected to encompass an intermediate temperature range,
neither too close to $T_{c}$ nor to zero.  A significant feature of 
the concavity plots shown in Fig.~\ref{Fig:fig4} is weak dependence
of the concavity of the magnetization profiles on the carrier density;
though the three values of $n_{c}$ range over two orders of magnitude 
($10^{10}-10^{12})~\textrm{cm}^{-2}$, the $\gamma$ graphs lie very close to 
one another.  One can also see that the parameter range over 
which the $M(T)$ profile is concave is much more extensive than 
predicted by lattice MFT for three dimensional DMS
systems~\cite{tres}.  This is primarily a consequence of the geometry 
of the two dimensional lattice.  The inset of  
Fig.~\ref{Fig:fig4} displays representative cases of the magnetization 
$M(T)$.  

     In conclusion, we have considered diluted magnetic semiconductors 
in two dimensions.  We find that long-range 
ferromagnetic order can be stabilized by even a small amount of 
anisotropy invariably present in reality.  We have found 
that, even at the level of continuum
MFT, finite temperature effects in the carrier-mediated
effective interaction between Mn impurity moments introduce a strong 
dependence on the density of carriers, where naive use of the zero 
temperature RKKY formula yields a result independent of $n_{c}$.
To take into account the discreteness of the strong positional disorder
of the 2D DMS system, we have employed a lattice MFT.  
Our lattice theory also provides a convenient framework for the 
inclusion of important physics such as the finite mean free path.
In general, the 2D DMS $T_{c}$ is strongly suppressed compared with
the corresponding 3D DMS $T_{c}$, which is somewhat discouraging for 
spintronic applications.

This work is supported by US-ONR and DARPA.

\end{document}